\documentclass[11pt, oneside]{article}      
\usepackage{geometry}                		
\usepackage[parfill]{parskip}    		         
\usepackage{graphicx}				 
\usepackage{amssymb}
\usepackage[T1]{fontenc}
\usepackage{authblk}
\usepackage[utf8]{inputenc}
\usepackage{xcolor}
\usepackage{amsmath}
\usepackage{algorithm}
\usepackage{algorithmicx}
\usepackage{algpseudocode}
\usepackage{mathtools}

\title{Frequently Co-cited Publications: Features and Kinetics }

\author[1]{Sitaram Devarakonda}
\author[2]{James Bradley}
\author[1]{Dmitriy Korobskiy}
\author[3]{Tandy Warnow}
\author[1]{George Chacko\thanks{netelabs@nete.com}}

\affil[1]{Netelabs, NET ESolutions Corporation, McLean, VA}
\affil[2]{Raymond Mason School of Business, Coll. of William \& Mary, Williamsburg, VA}
\affil[3]{Department of Computer Science, Univ. of Illinois, Urbana-Champaign, IL}

\begin{document}
\maketitle
\newpage

\begin{abstract} Co-citation measurements can reveal the extent to which a concept representing a novel combination of existing ideas evolves towards a specialty. The strength of co-citation is 
represented by its frequency, which accumulates over time. Of interest is whether underlying features associated with the strength of co-citation can be identified. 
We use the proximal citation network for a given pair of articles $(x, y)$ to compute $\theta$, an \emph{a priori} estimate of the probability of co-citation between x and y, prior to their first co-citation.Thus, 
low values for $\theta$ reflect pairs of articles for which co-citation is presumed less likely. We observe that co-citation frequencies are a composite of power-law 
and lognormal distributions, and that very high co-citation frequencies are more likely to be composed of pairs with low values of $\theta$, reflecting the impact of a novel combination of ideas. Furthermore, we note 
that the occurrence of a direct citation between two members of a co-cited pair increases with co-citation frequency. Finally, we identify cases of frequently co-cited publications that accumulate co-citations after an 
extended period of dormancy.\end{abstract}

\section*{Introduction} Co-citation, ``the frequency with which two documents from the earlier literature are cited together in the later literature'', was first described in 
1973~\cite{marshakova-shaikevich_system_1973,small_co_citation_1973}. As noted by \cite{small_co_citation_1973}, 
co-citation patterns differ from bibliographic coupling patterns~\cite{kessler_bibliographic_1963} but align  with patterns of direct citation and  
frequently co-cited publications must have high individual citations.

Co-citation has been the subject of further study and characterization, for example, comparisons to bibliographic coupling and direct 
citation~\cite{boyack_cocitation_2010}, the study of invisible colleges~\cite{gmur_co-citation_2003,noma_co-citation_1984}, 
construction of networks by co-citation~\cite{small_clustering_1985,small_clustering_1985-1}, evaluation of clusters in combination with 
textual analysis~\cite{braam_mapping_1991}, textual similarity at the article and other levels~\cite{colavizza_closer_2018}, 
and the fractal nature of publications aggregated by co-citations~\cite{vanraan_fractal_1990}.

Co-citations provide details of the relationship between key (highly cited) ideas, and changes in co-citation patterns over time may provide insight into 
the mechanism with which new schools of thought develop. Implicit in the definition of co-citation is novel combinations of existing ideas, but
only some frequently co-cited article pairs reflect surprising combinations. For example, two publications presenting the leading methods for the same 
computational problem may be highly co-cited, but this does not reflect a novel combination of ideas. Similarly, two publications describing methods that often constitute
part of the same workflow may be highly co-cited, but these co-citations are also not surprising.
On the other hand, for two articles in different fields, frequent co-citation is generally unexpected.

Novel, atypical, or otherwise unusual combinations of co-cited articles have been explored at the 
journal-level~\cite{wang_2017,bradley_co-citations_2019,boyack_vs_uzzi_2014,uzzi_atypical_2013}. However,  
journal-level classifications have limited resolution relative to article-level studies, which may better represent the actual structure and aggregations of the scientific 
literature~\cite{shu_comparing_2019,article_boyack_topic,waltman_new_2012,article_stasa,article_gomez_journal}.  Accordingly, we sought to discover measurable 
characteristics of frequently co-cited publications from an article-level perspective.

To study frequently co-cited articles, we have developed a novel graph-theoretic approach that reflects the citation 
neighborhood of a given pair of articles. In seeking to determine the degree to which a co-cited pair of papers represented a surprising combination, 
we wished to avoid journal-based field classifications, which present challenges.
Instead, we attempted to use citation history to produce an estimate of the 
probability that a given pair of publications $(x, y)$ would be co-cited. Since we focus on the activity before they are first co-cited, the 
``probability" of co-citation is zero, by definition, since there are no co-citations yet. Hence, we approximated co-citation probabilities: we treat an article that cites one member 
of a co-cited pair and also cites at least one article that cites the other member as a proxy for co-citation. Specifically, given a pair of publications $x,y$, we construct a directed  bipartite graph 
whose vertex set contains  all publications that cite either $x$ or $y$ previous to their first co-citation. We then compute $\theta$, a normalized count of such 
proxies, and use it to predict the probability of co-citation between $x$ and $y$. This approach enables an evaluation that is specific to the given pair of articles, 
and does so without substantial computational cost, while avoiding definitions of disciplines derived 
from journals or having to measure disciplinary distances.

To support our analysis, we constructed a dataset  of articles from Scopus~\cite{scopus_ref} that were published in the eleven year period, 1985-1995, and 
extracted the cited references in these articles. Recognizing that frequently co-cited publications must derive from highly-cited publications~\cite{small_co_citation_1973}, 
we identified those reference pairs (33.6 million pairs) for each article in the dataset that are drawn from the  top 1\% most cited articles in Scopus and measured their frequency 
of co-citation.

To investigate which statistical distributions might best describe the co-citation frequencies in these 33.6 million co-cited pairs, we reviewed prior work on distributions of citation frequency~\cite{radicchi_statistical_2008,eom_2011,price_1965,price_general_1976,newman_structure_2003,wang_quantifying_2013,stringer_statistical_2008,stringer_statistical_2010,redner_statistical_2005}. 
This research has fit the frequency distribution of citation strength sometimes to a power law distribution and other times to a lognormal distribution.  
A graph of the analogous co-citation data suggests that power law or lognormal distributions are candidates for describing co-citation strength as well and so we, accordingly, 
investigated that conjecture.  Interestingly, \cite{Mitzenmacher_2003} notes the debate between the appropriateness of power law versus lognormal distributions is 
not confined to bibliometrics, but has been at issue in many disciplines and contexts.

To study how the best-fit distributional function and parameters for co-citation might vary with $\theta$, we stratified co-citation frequency data. We also measured whether a direct link exists between 
two members of a co-cited pair (i.e., whether one member of a pair cites the other) and how this property is related to co-citation frequencies. 
We find that the distribution of co-citation frequencies varies with $\theta$ and that a
power law distribution fits co-citation frequencies more often when $\theta$ is small, whereas a lognormal distribution fits more often for large $\theta$.

A pertinent aspect of co-citation is the rate at which frequencies accumulate. While citation dynamics of individual publications have been fairly well studied 
by others, for example, \cite{wallace_2009,eom_2011}, the dynamics of co-cited articles are less well studied. Our interest was the special case analogous to the Sleeping Beauty 
phenomenon~\cite{vanraansleeping2004,ke_defining_2015}, which may reflect delayed recognition of scientific discovery and the causes attributed to 
it~\cite{merton_1963,garfield_1970,garfield_1980,cole_1970,barber_1961,glanzel_myth_2004}. Thus, we also identified co-cited pairs that 
featured a period of dormancy before accumulating co-citations.

\section*{Materials \& Methods} \emph{Data} Citation counts were computed for all Scopus articles (88,639,980 records) updated through December 2019, as implemented in the 
ERNIE project~\cite{GithubERNIE2019}. Records with corrupted or missing publication years or classified as `dummy' by the vendor were then removed, 
resulting in a dataset of 76,572,284 publications. Hazen percentiles of citation counts, grouped by year of publication, were calculated for the these 
data~\cite{bornmann_use_2013}. The top 1\% of highly cited publications from each year were combined into a set of highly cited publications consisting of 768,993 publications.

Publications of type `article', each containing at least five cited references and published in the 11 year period from 1985-1995, were subset from Scopus to form a dataset 
of 3,394,799 publications and  51,801,106 references (8,397,935 unique). For each of these publications, all possible  reference pairs were generated and then restricted to 
those pairs where both members were in the set of highly cited publications (above). 

For example, the data for 1985 consisted of 223,485 articles after processing as described above. Computing all reference pairs (that were also members of the highly cited 
publication set of 768,993) from these  223,485 articles gave rise to 2,600,101 reference pairs (Table~\ref{tab:tab1}) that ranged in co-citation frequency from 1 to 874 within the 1985 
dataset; from 1 to 11,949 across the 11 year period 1985-1995; and from 1 to 35,755 across all of Scopus. Collectively, the publications in our 1985-1995 dataset generated 33,641,395 
unique co-citation pairs, for which we computed co-citation frequencies across all of Scopus.

\begin{figure}[ht] 
\centering
\includegraphics[width=0.7\textwidth]{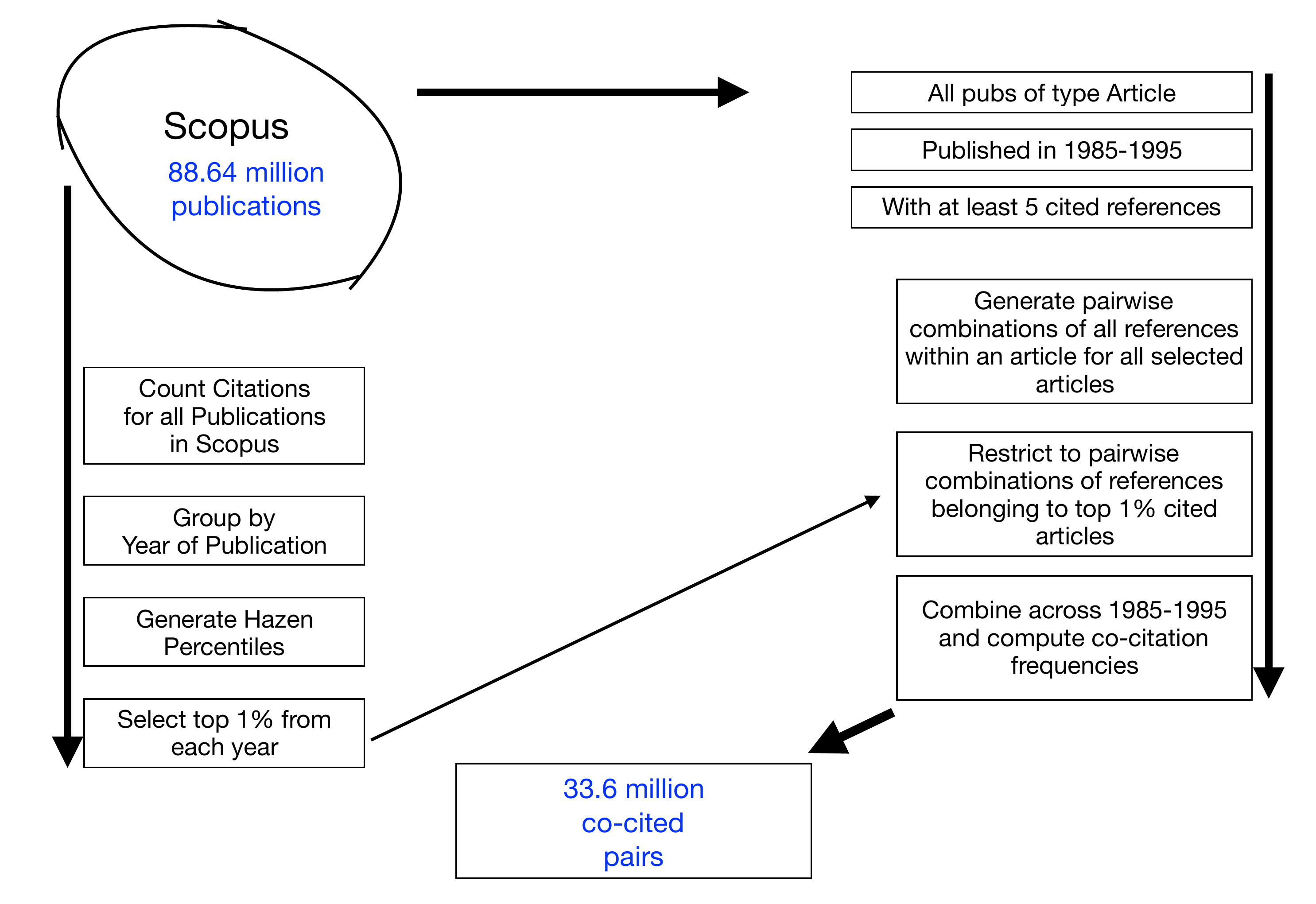}
\caption{\bf The workflow we used to generate a dataset of  33,641,395 co-cited publications from references cited by articles in Scopus published in the years 1985-1995.}
\label{fig:schematic}
\end{figure}

\begin{table}[!ht]
\caption{{\bf Summary of Analyzed Data} Publication of type article that had at least five cited references indexed in Scopus were selected from the eleven years, 1985-1995. All possible reference pairs  
were generated for the cited references of these articles and then restricted to those pairs where both members  were in the set of 768,993 highly cited publications. The column Co-cited Pairs shows 
the number of pairs in each year after the restriction was applied.} 
\label{tab:tab1}
\centering
\begin{tabular}{lllc}
\hline
Year  & Articles & References  & Co-cited Pairs\\
\hline
1985 & 223,485   & 1,796,502 & 2,600,101 \\
1986 & 238,096   & 1,920,225 & 2,840,557 \\
1987 & 250,575   & 2,037,654 & 3,180,261 \\
1988 & 269,219   & 2,182,571 & 3,406,902 \\ 
1989 & 285,873   & 2,303,481 & 3,793,986 \\
1990 & 305,010   & 2,490,909 & 4,546,915 \\
1991 & 325,782.  & 2,662,005 & 5,039,334 \\
1992 & 343,239.  & 2,846,607 & 5,622,164 \\
1993 & 360,916   & 3,006,374 & 6,121,147 \\
1994 & 387,062.  & 3,228,240 & 7,022,499 \\
1995 & 405,503.  & 3,432,228 & 7,626,684 \\
\hline
\end{tabular}
\end{table}

\emph{Derivation of $\theta$ } 
We now show how we define our prior on the probability of $x$ and $y$ being co-cited, based on the  citation graph restricted to publications that cite either $x$ or $y$ (but not both) up to the year of their
first co-citation. 
Recall that we defined  a proxy co-citation of $x$ and $y$ to be
an article that cites one member of the co-cited pair $(x,y)$ and also cites at least one article that cites the other member.
The idea behind this definition is that
we consider papers  that cite $x$ as proxies for $x$, and papers  that cite $y$ as proxies for $y$. Thus, if a paper $a$   cites both $x$ and $y'$ (where $y'$ is a proxy for $y$), then it is a proxy for a 
co-citation of $x$ and $y$. Similarly, if a paper $b$  cites both $y$ and $x'$ (where $x'$ is a proxy for $x$), it is also a proxy for a co-citation of $x$ and $y $.
This motivates the graph-theoretic formulation, which we now formally present.

We  fix the pair $x, y$ and we define  $N(x)$ to be the  set of all   publications that cite $x$ (but do not also cite $y$), and are published no later than the year of the first co-citation of $x$ and $y$.
We similarly define $N(y)$.  We define a directed bipartite graph  with vertex set $N(x) \cup N(y)$. Note that if $x$ cites $y$ then $x \in N(y)$, and similarly for the case where $y$ cites $x$.
Note also that since we have restricted $N(x)$ and $N(y)$ that  $N(x) \cap N(y) = \emptyset$. We now describe how the directed edge set $E(x,y)$ is constructed. 
For any pair of articles $a,b$ where $a \in N(x)$ and $b \in N(y)$, if $a$ cites $b$ then we include the directed edge $a \rightarrow b$ in $E(x,y)$.
Similarly, we include edge $b \rightarrow a$ if $b$ cites $a$. Finally,  if a pair of articles both cite each other, then the graph has parallel edges.
By construction, this graph is {\em bipartite}, which means that all the edges go between the two sets $N(x)$ and $N(y)$ (i.e., no edges exist between two vertices in $N(x)$, nor between two vertices
in $N(y)$).

Note that by the definition,  every edge in $E(x,y)$ arises because of a proxy co-citation, so that the number of proxy co-citations  is the number of directed edges in $E(x,y)$. 
Consider the situation where a publication $a$  cites $x$ (so that $a \in N(x)$) and also cites $b_1, b_2, b_3$  in $N(y)$: this defines three directed edges from $a$ to nodes of $N(y)$.
We count this as three proxy co-citations, not as one proxy co-citation. Similarly, if we have a publication $b$ that cites $y$ and also  cites $a_1, a_2, a_3, a_4$ in $N(x)$, 
then there are four directed edges that go from $b$ to nodes in $N(x)$ and we will count each of those  directed edges as a different proxy co-citation.

Accordingly, letting $|X|$ denote the cardinality of a set $X$,  we note $|E(x,y)|$, i.e., the number of directed edges that go between $N(x)$ and 
$N(y)$, is the number of proxy co-citations between $x$ and $y$.
If no parallel edges are permitted, the maximum number of   possible proxy co-citations is $|N(x)| \times |N(y)|$.
Under the assumption that both $N(x)$ and $N(y)$ each have at least one article, we 
define $\theta(x,y)$,  our prior on the probability of $x$ and $y$ being co-cited, as follows:
$$\theta(x,y)= \frac{|E(x,y)|}{ |N(x)| \times |N(y)|} . $$ 

Note that if parallel edges do not occur in the graph, then $\theta(x,y) \leq 1$, but that otherwise the value can be greater than $1$.
Note also that $\theta(x,y)=0$ if $E(x,y) = \emptyset$ (i.e., if there are no proxy co-citations) and
that $\theta(x,y)=1$ if every possible proxy co-citation occurs.

To efficiently calculate $\theta$, we used the following pipeline. We copied Scopus data from a relational schema in PostgreSQL into 
a citation graph from Scopus into the Neo4j 3.5 graph database using an automated Extract Transform Load (ETL) pipeline that combined
Postgres CSV export and the Neo4j Bulk Import tool. The graph vertex set is all publications, each with a publication year attribute, and 
the edge set is all citations between the publications. A Cypher index was created on the publication year. We developed Cypher queries 
to calculate $\theta$ and tuned performance by splitting input publication pairs into small batches and processing them in parallel, using parallelization 
in Bash and GNU Parallel. Batch size, the number of parallel job slots, and other parameters were tuned for performance, with best results achieved on 
batch sizes varying from 20 to 100 pairs. The results of $\theta$ calculations were cross-checked using SQL calculations. In the small number of cases where  
$\theta$ computed to $> 1$ (above) it was set to 1 for the purpose of this study.

\emph{Statistical Calculations} We denote the observed co-citation frequency data by the multi-set
\begin{displaymath}
X^o = \left\lbrace x^o_1, \ldots , x^o_N  \right\rbrace ,
\end{displaymath}
where  $N$ is the total number of pairs of articles and $x^o_i$ is the observed frequency of the $i^{th}$ pair of papers being co-cited. 
Note that this is in general a multi-set, as different pairs of articles can have the same co-citation frequency.
Let $n(x)$ be the number of times that $x$ appears in $X^o$ (equivalent, $n(x)$ is the number of pairs of articles that 
are co-cited $x$ times), and let $N(x) = \sum_{y=x}^{\infty}{n \left( y \right)}$ denote the total number of pairs of articles that are co-cited at least $x$ times.
Then 
\begin{equation} \label{eqObsFreqDist}
  f^o \left( x \left| x \geq \underline{x} \right. \right) = \frac{ n(x)}{N(\underline{x})}  \text{   for } x \in \left\lbrack \underline{x} , \infty \right) ,
\end{equation}
where $\underline{x}$ is a parameter we use to analyze the distribution's right tail starting at varying frequencies. We describe in this subsection (i) the statistical computations for fitting lognormal and power law distributions to right tails of the observed co-citation frequency distributions as defined by (\ref{eqObsFreqDist}) for various $\underline{x}$ and (ii) how we assessed the quality of those fits. Further, we performed such analyses for various slices of the data, stratifying by $\theta$ and other parameters, as is described in the Results section.

We used a discrete version of a lognormal distribution to represent integer co-citation frequencies, $f \left( \cdot \right)$, following ~\cite{stringer_statistical_2008} and ~\cite{stringer_statistical_2010}, while appropriately normalizing for our conditional assessment of the right tail commencing at $\underline{x}$:
\begin{eqnarray}
f_{LN} \left( x \left| \mu, \sigma, \underline{x} \right. \right)  &=& \frac{\tilde{f} \left( x \left| \mu, \sigma \right. \right) }{\sum_{n = \underline{x}}^\infty{\tilde{f} \left( n \left| \mu, \sigma \right. \right)}} \text{   for }  x \geq \underline{x} \label{eqLognormDist} \\
\tilde{f} \left( x \left| \mu, \sigma \right. \right)&=& \int_{ x - 0.5 }^{ x + 0.5 } \frac{dq}{q \sqrt{2 \pi \sigma^2}} \exp \left( -\frac{\left( \ln{q} - \mu  \right)^2}{2 \sigma^2} \right)  , \nonumber
\end{eqnarray}
where $\mu$ and $\sigma$ are the mean and standard deviation, respectively, of the underlying normal distribution.  These probabilities can be computed with the cumulative normal distribution,
\begin{displaymath}
\tilde{f} \left( x \left| \mu, \sigma \right. \right) = \Phi \left( \frac{\ln{\left( x + 0.5 \right)}}{\sigma} \right)  
         -  \Phi \left( \frac{\ln{\left( x - 0.5 \right)}}{\sigma} \right) ,
\end{displaymath}
using the well-known error function.

We fit distributions to the co-citation frequency data for various extremities of the right tail, as parameterized by $\underline{x}$, using a maximum (log) likelihood estimator (MLE). We solved for the best-fit distributional parameters for the lognormal distribution, $\mu$ and $\sigma$, by modifying a multi-dimensional interval search algorithm from \cite{press_statistical_2007} and following ~\cite{stringer_statistical_2010}. A compiled version of this code using the C++ header file, \texttt{amoeba.h}, is available on our Github site~\cite{GithubERNIE2019}.  

We fit a discrete power law distribution to the data for various values of $\underline{x}$, which was normalized for our conditional observations of the right tail:
\begin{equation}
f_{PL} \left( x \left| \alpha , \underline{x}  \right. \right) = \frac{x^{-\alpha}}{\zeta\left( \alpha , \underline{x} \right)} \text{   for } x \geq \underline{x},
\end{equation}
where the Hurwitz zeta function, 
\begin{displaymath}
\zeta \left( \alpha , \underline{x} \right) = \sum_{x=0}^{\infty} {\frac{1}{\left( x + \underline{x}  \right)^{\alpha}}} ,
\end{displaymath}
is a generalization of the Riemann zeta function, $\zeta \left( \alpha , 1 \right)$, as is needed for analysis of the right tail.

We solved first-order conditions for the (log) MLE to find the best-fit distributional exponent $\alpha$,
\begin{equation} \label{eqFirstOrderPL}
\frac{\zeta^\prime \left( \alpha , \underline{x} \right)}{\zeta \left(  \alpha , \underline{x} \right)} = - \frac{1}{N \left( \underline{x} \right)} \sum_{x \in X^o \left(  \underline{x} \right)} \ln{x}  ,
\end{equation}
as described in \cite{clauset_power-law_2009} and ~\cite{goldstein_problems_2004}, where $X^o \left(  \underline{x} \right) = \left\lbrace x \in X^o : x \geq \underline{x} \right\rbrace$,  are the observed co-citations with frequencies at least as great as $\underline{x}$ and $N \left( \underline{x} \right)$ is the number of such co-citations. 
 We solved (\ref{eqFirstOrderPL}) to find $\alpha$  using a bisection algorithm.

We used the ${\chi}^2$ goodness of fit (${\chi}^2$) and the Kolmogorov-Smirnov (K-S) tests to assess the null hypothesis that the distribution of the observed co-citation frequencies and the best-fit lognormal distribution are the same, and similarly for the best-fit power law distribution. We also computed the Kullback-Leibler Divergence (K-L) between the observed data and the best-fit distributions.

Both the ${\chi}^2$ and K-S tests employed the null hypothesis that the observed co-citation frequencies, $n{\left( x \right)}$ for $ x \in \left\lbrack \underline{x} , \infty \right)$, were sampled from the best-fit lognormal or power law distributions, which we denote by  $f_d \left( \cdot \left| \underline{x} \right. \right)$ for $d \in \left\lbrace LN, PL \right\rbrace$, while suppressing the parameters specific to each of the distributions.  

The usual ${\chi}^2$ statistic was computed by, first, grouping each of the observed co-citation frequencies into $k$ bins, denoted by $b_i$ for $i \in \left\lbrace 1, \ldots , k \right\rbrace$, and then computing
\begin{displaymath}
  \chi^2 = \sum_{i=1}^{k} {\frac{\left( O_i - E_i \right)^2}{E_i}},
\end{displaymath}
where $O_i$ is the observed number of co-citations having frequencies associated with the $i$-th bin,
\begin{displaymath}
  O_i = \sum_{x \in b_i}{n{\left( x \right)}} ,
\end{displaymath}
and $E_i$ is the expected number of observations for frequencies in bin $i$, if the null hypothesis was true, in a sample with size equal to the number of observed data points, $N{\left( \underline{x} \right)}$:
\begin{displaymath}
  E_i = \sum_{x \in b_i} f_d \left( x \left| \underline{x}  \right. \right) N{\left( \underline{x} \right)}
\end{displaymath}
If the null hypothesis was true, then we would expect $O_i$ and $E_i$ to be approximately equal, with deviations owing to variability due to sampling.

Constructing the bins $b_i$ requires only that $E_i \geq 5$ for every $i = 1, \ldots , k$.    Test outcomes are sometimes sensitive to the minimum $E_i$ permitted, which we will denote by $\underline{E}$, and so we tested with multiple thresholds, including 10, 20, 50, and 70. Furthermore, statistical tests are stochastic: these multiple tests permitted a reduction in the probability of erroneously rejecting or accepting the null hypothesis based on a single test.  The distribution of observed co-citation frequencies was skewed right with a long tail, so that aggregating bins to satisfy $E_i \geq \underline{E}$ was most critical in the right tail. This motivated a bin construction algorithm that aggregated frequencies in reverse order, starting with the extreme right tail. Algorithm \ref{alg:bins} requires a set of the unique observed co-citation frequencies, $\hat{X}^o$, which includes the elements of the multiset $X^o$ without repetition.  While Algorithm \ref{alg:bins} does not guarantee in general that all bins satisfy $E_i \geq \underline{E}$, that criterion was satisfied for the observed data. 

\begin{algorithm}
  \caption{Frequency Bin Construction
    \label{alg:bins}}
  \begin{algorithmic}[1]
    \State $i \leftarrow 1$
    \State $b_1 = \left\lbrace \right\rbrace$
    \While {$\left| \hat{X}^o \right| > 0$}
      \State $b_i \leftarrow b_i \cup \left\lbrace \max{\left( \hat{X}^o \right)} \right\rbrace$
      \State $\hat{X}^o \leftarrow \hat{X}^o \setminus \max{\left( \hat{X}^o \right)}$
      \If {$E_i \geq \underline{E}$}
        \State $i \leftarrow i + 1$
        \State $b_i \leftarrow \left\lbrace \right\rbrace$
      \EndIf
    \EndWhile
  \end{algorithmic}
\end{algorithm}

We implemented a K-S test using simulation to generate a sampling distribution to account for the discrete frequency observations \cite{internetks_bootstrap}. We denote the cumulative 
distribution of observed co-citation frequencies by $F^o{\left( x \left| \underline{x} \right. \right)} = \sum_{i=\underline{x}}^{x} {f^o{\left( i \left| \underline{x} \right. \right)}}$, and the best-fit cumulative 
distribution by $F_d{\left( x \left| \underline{x} \right. \right)} = \sum_{i=\underline{x}}^{x} {f_d{\left( i \left| \underline{x} \right. \right)}}$.  The K-S test involves testing the maximum absolute difference 
between the observed and theorized cumulative distributions, 
\begin{displaymath}
  D_n  = \max_x{\left| F^o{\left( x \left| \underline{x} \right. \right)} - F_d{\left( x \left| \underline{x} \right. \right)}  \right|},
\end{displaymath}
where $n$ is the number of observations giving rise to $F^o{\left( x \left| \underline{x} \right. \right)}$, against the distribution of such differences between samples from the theorized distribution with the same number of observations, $n$,
\begin{displaymath}
  \tilde{D}_n  = \max_x{\left| \tilde{F}_{d,1}{\left( x \left| \underline{x} \right. \right)} - \tilde{F}_{d,2}{\left( x \left| \underline{x} \right. \right)}  \right|} ,
\end{displaymath}
where $\tilde{F}_{d,j}{\left( x \left| \underline{x} \right. \right)}$ is the empirical distribution of sample $j$ of size $n$ (notation suppressed) drawn from $F_d{\left( x \left| \underline{x} \right. \right)}$.  We generated 100 such random variables $\tilde{D}_n$ for each test.  We reject the null hypothesis if $D_n$ is larger than substantially all of the $\tilde{D}_n$, say all but 5\%, for equivalence with a $p$-value of 0.05. The number of $\tilde{D}_n$ samples drawn yields a $p$-value with a resolution of 1\%.

We computed the K-L Divergence two ways due to its asymmetry:
\begin{eqnarray*}
D_{K-L}{\left( f^o \parallel f_d \right)} &=& \sum_{x=\underline{x}}^\infty {f^o{\left( x \left| \underline{x} \right. \right)}  \ln{\frac{f^o{\left( x \left| \underline{x} \right. \right)}}{f_d{\left( x \left| \underline{x} \right. \right)}}}}  \\
D_{K-L}{\left( f_d \parallel f^o  \right)} &=& \sum_{x=\underline{x}}^\infty {f_d{\left( x \left| \underline{x} \right. \right)}  \ln{\frac{f_d{\left( x \left| \underline{x} \right. \right)}}{f^o{\left( x \left| \underline{x} \right. \right)}}}} .
\end{eqnarray*}

Separate from the tests above, we tested whether the distribution of co-citation frequencies was independent of $\theta$ using a $\chi^2$ test, using the null hypothesis that the co-citation frequency distribution was 
independent of $\theta$ . We initially created a contingency table on $\theta$ and co-citation frequency using these bins for $\theta$, $\left\lbrace \left\lbrack 0.0 , 0.2 \right) ,  \left\lbrack 0.2 , 0.4 \right) ,   \left\lbrack 0.4 , 0.6 \right) , 
\left\lbrack 0.6 , 0.8 \right) ,  \left\lbrack 0.8 , 1.0 \right)  \right\rbrace$, and logarithmic bins for 
frequency to accommodate the skewed distributions: 
\begin{displaymath}
\left\lbrace \left[ 10 , 100 \right) , \left[ {100} , {1000} \right) ,  \left[ {1000} , {10000} \right) ,  \left[ {10000} , {100000} \right]  \right\rbrace.
\end{displaymath}
We, subsequently, aggregated these bins to have an expected number of co-citations in each bin equal to or greater than 5 to account for a decreasing number of observations as $\theta$ and frequency increased by having just two 
intervals for frequency: $\left\lbrace \left[ 10 , 100 \right) , \left[ {100} ,  {100000} \right]  \right\rbrace$.

\emph{Kinetics of Co-citation} 
We extended prior work on delayed recognition and the Sleeping Beauty phenomemon~\cite{ke_defining_2015,vanraansleeping2004,li_distinguishing_2016,glanzel_myth_2004} towards co-citation. 
We have modified the beauty coefficient (B) of ~\cite{ke_defining_2015}  to address co-citations by:
 (i) counting citations to a pair of publications (co-citations) rather than citations to individual papers,
  (ii) setting $t_0$ (age zero) to the first year in which a pair of publications could be co-cited (i.e., the publication year of the more recently published member of a co-cited pair), and
   (iii)  setting $C_0$ to the number of co-citations occurring in year $t_0$. Rather than calculate awakening time as in \cite{ke_defining_2015}, we opted to measure the simpler length of time between $t_0$ and the first year in which a co-citation was recorded; we label this measurement as the timelag $t_l$, so that $t_l=0$  if a co-citation was recorded in $t_0$.

\section*{Results and Discussion} 

\begin{figure}[h] 
\centering
\includegraphics[width=0.7\textwidth]{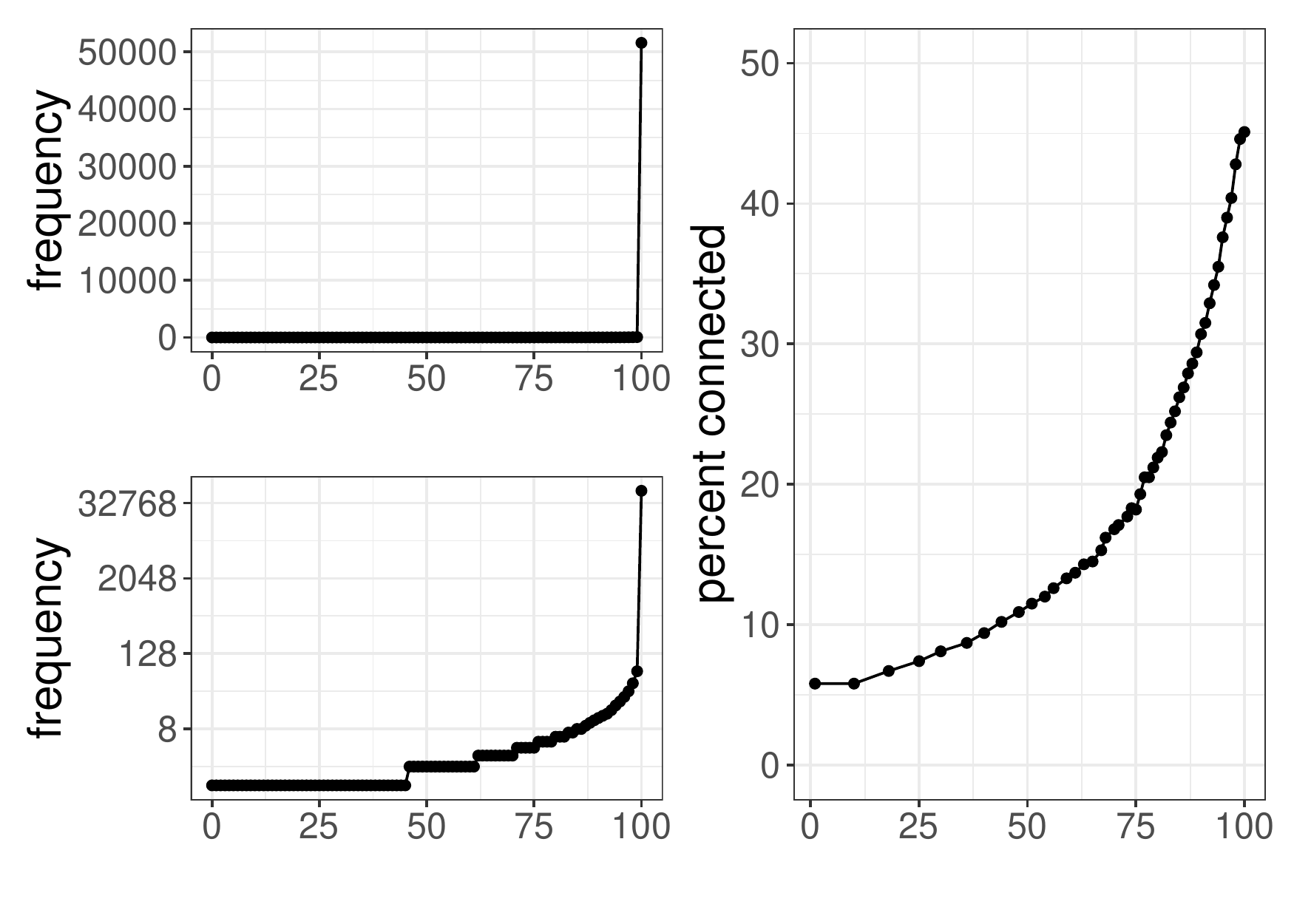}
\caption{The x-axis shows percentiles for all three plots \textbf{Left Side} \emph{Co-citation frequencies of highly cited publications from Scopus 1985-1995} Co-citation frequencies are plotted against their percentile values.
The upper and lower plots were both generated from 33,641,395 data points. The lower plot shows the same data with a logarithmic (ln) transformation of y-axis. The minimum co-citation frequency is 1, the median is 2, the third 
quartile is 4, and the maximum is 51,567. Additionally, 15,140,356 pairs (45 \%) have a co-citation frequency of 1. Frequencies of 12, 22, 67, and 209 correspond to quantile values of 0.9, 0.95, 0.99, and 0.999 respectively.
\textbf{Right Side} Direct citations between members of a co-cited pair (connectedness) increase with co-citation frequency. The proportion of connected pairs (a direct citation exists between the two members of a pair) within each 
percentile is shown. Data are plotted for all pairs with a co-citation frequency of at least $10$
(4.1 million pairs)}
\label{fig:basedata}
\end{figure}

Our base dataset, described in  Table \ref{tab:tab1}, consists of the 33,641,395 co-cited reference pairs (33.6 million pairs) and their co-citation frequencies, gathered from Scopus during the 11-year period from 1985-1995 (Materials and Methods). A striking distribution of co-citation frequencies with a long right tail is observed with a minimum co-citation of 1, a median of 2, and a maximum co-citation frequency of 51,567 (Figure~\ref{fig:basedata}).  Approximately 33.3 of 33.6 million pairs (99\% of observations) have co-citation frequencies ranging from 1--67 and the remaining 1\% have co-citation frequencies ranging from 68--51,567.  Since the focus of our study was co-citations of frequently cited publications, we further restricted this dataset to those pairs with a co-citation frequency of at least 10, which resulted in a smaller dataset of 4,119,324 co-cited pairs (4.1 million pairs) with minimum co-citation frequency of 10, median of 18, and a maximum co-citation frequency of 51,567. In order to focus on co-citations derived from highly cited publications, $\theta$ was calculated for all pairs with a co-citation frequency of at least 10. We also note whether one article in a co-citation pair cites the other (connectedness).

Influenced by the use of linked co-citations for clustering~\cite{small_clustering_1985}, we also examined the extent to which members of a co-cited pair were also found in other co-cited pairs. We found that 205,543 articles contributed to 4.12 million co-cited pairs. The highest frequency observed in our dataset, 51,567 co-citations, was for a pair of articles from the field of physical chemistry: Becke (1993)~\cite{becke_densityfunctional_1993} and Lee, Yang, and Parr (1988)~\cite{lee_development_1988}. The members of this pair are not connected and are found in a total of 1,504  co-cited pairs with frequencies ranging from 10 to 51,567.  The second highest frequency, 28,407 co-citations, was for  another pair of articles from the field of 
biochemistry: \cite{laemmli_cleavage_1970,bradford_rapid_1976}. Members of this pair are not connected and are found in 41,909 co-cited pairs, 24,558 for the Laemmli gel electrophoresis article and 17,352 for the Bradford protein estimation article. In terms of this second pair, both articles describe methods heavily used in biochemistry and molecular biology, an area with strong referencing activity, so this result is not entirely surprising. 

Having developed $\theta(x,y)$ as a prediction of the probability that   articles $x$ and $y$ would be co-cited, we first tested whether the distribution of co-citation frequencies was independent of $\theta$ (Materials and Methods). The null hypothesis that the co-citation frequency distribution was independent of $\theta$ was rejected with a very small $p$-value: the statistical software indicated a $p$-value with no significant non-zero digits.  We next investigated what distribution functions might fit the frequencies of co-citation as $\theta$ varied.

\begin{figure}[h]
\centering
\includegraphics[width=0.6\textwidth]{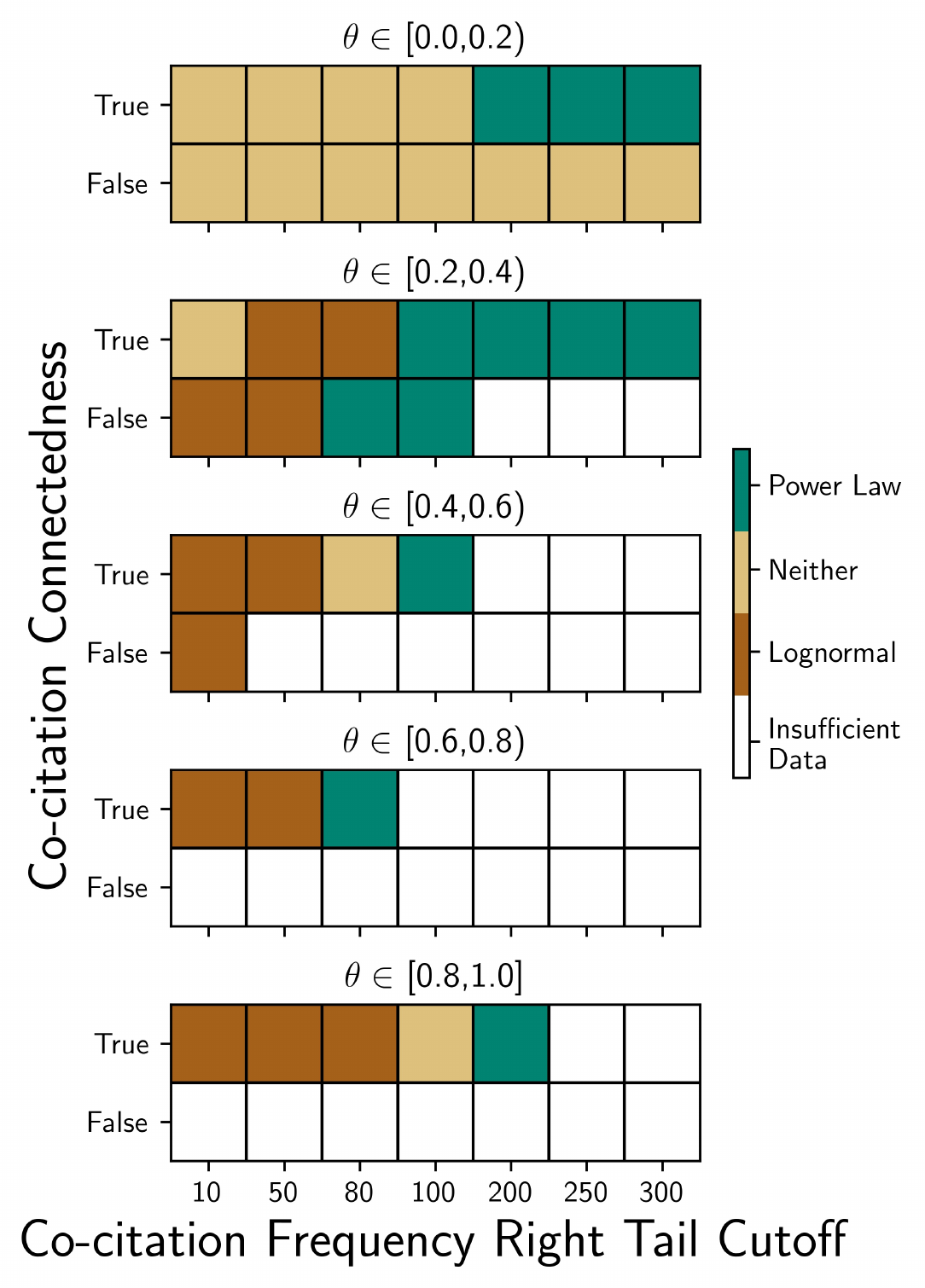}
\caption{\textbf{Distributional fits to the observed co-citation frequencies} The graph shows where a lognormal or power law distribution demonstrated a statistically significant fit with the observed co-citation frequencies 
stratified by $\theta$, extent of the right tail tested $\underline{x}$, and whether co-citations were connected.  A power law fit more often for $\theta$ in the intervals $\left[ 0.0, 0.2 \right)$ and $\left[ 0.2, 0.4 \right)$ when 
cocitation constituents were connected.  When a lognormal distribution fit, it was for broader portions of the data set.  Data were insufficient for testing as $\theta$ increased due to (i) fewer observations and (ii) less prominent right tails.}
\label{fig:mondrian}
\end{figure}

Based on the long tails of citation frequencies, prior research has assessed the fit of lognormal and power law distributions \cite{stringer_statistical_2008,radicchi_statistical_2008,stringer_statistical_2010}. 
We noted long right tails in co-citation frequencies, which, similarly, motivated us to assess the fit of lognormal and power law distributions to co-citation data. Further, we stratified the data according to (i) the minimum frequency for the right tail $\underline{x}$, (ii) $\theta$, and (iii) whether the two members of each co-citation pair were connected.  Figure \ref{fig:mondrian} shows which distribution, if either, fits the data in each slice, based on tests of statistical significance. Note that there were no circumstances where both distributions fit: if one fit, then the other did not. 

\begin{figure*}
\begin{tabular}{cc}
\includegraphics[width=2.65in]{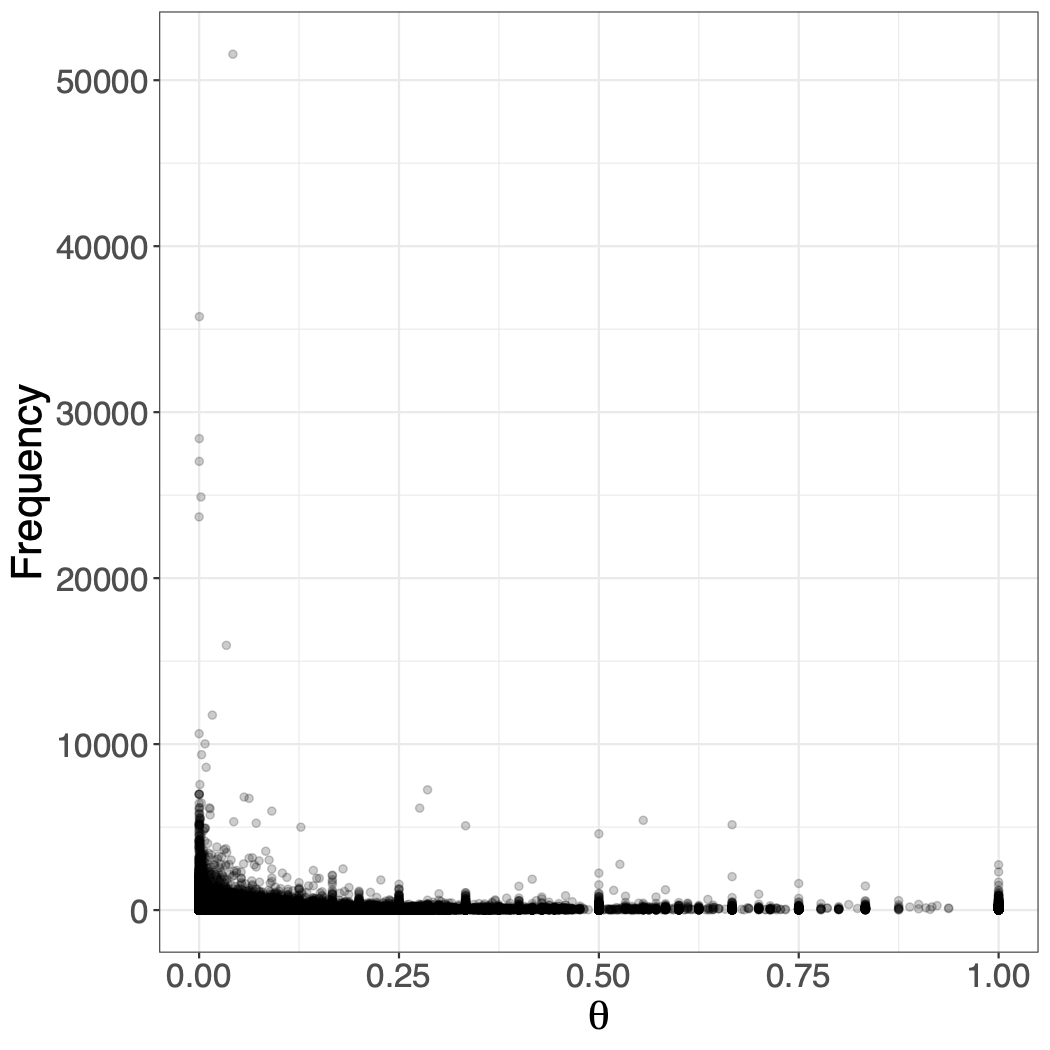} & 
\includegraphics[width=2.65in]{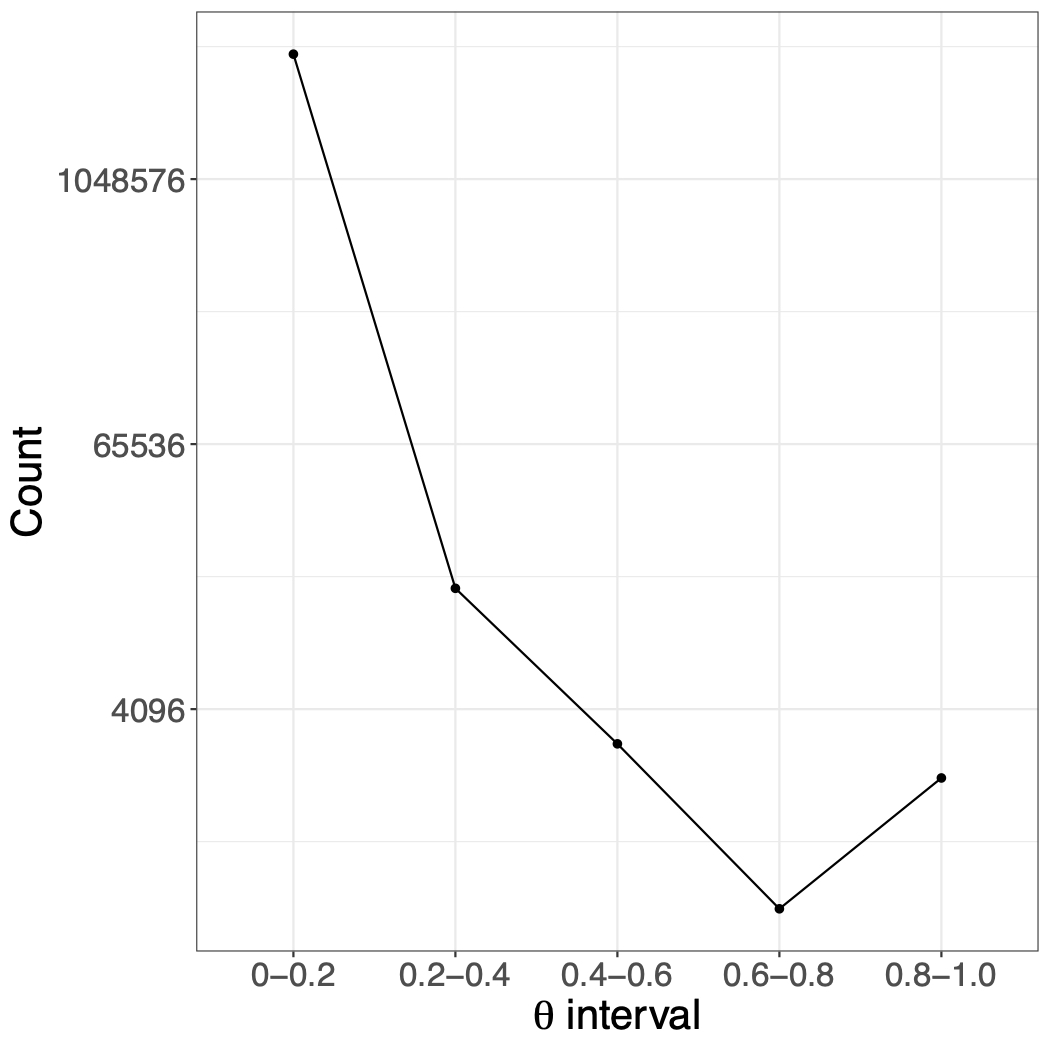}  \\
(a) Co-citation Scopus frequency versus $\theta$  & (b) Number of co-cited pairs per $\theta$ interval
\end{tabular}
\caption{Co-citation dynamics relative to  $\theta$.   (a) Points represent the Scopus frequency vs. $\theta$ value for  each co-cited pair.  Darker regions indicate denser plots of the translucent points.  
Co-cited pairs with the greater frequency are observed for pairs with smaller $\theta$.   (b) The $y$-axis employs a log scale and shows the number of co-cited pairs per $\theta$ interval. The number of co-cited pairs 
decreases, most often, by more than an order of magnitude per interval as $\theta$ increases. The dominance of co-cited pairs with  smaller $\theta$ are also reflected by regions of greater density in panel (a).}
\label{fig:Figs_4_5}
\end{figure*}

Statistical tests were not possible for some slices due to an insufficient number of data points. This was the case for certain combinations of large $\underline{x}$, large $\theta$, and co-citations that were not connected.   The number of data points obviously decreases as $\underline{x}$ increases, and we found the decrease in the number of data points to be more precipitous when $\theta$ was large and co-citations were unconnected due to the lighter right tails for these parameter combinations. The graph in the right panel of Figure \ref{fig:Figs_4_5}, which has a logarithmic $y$-axis, shows that the number of data points per $\theta$ interval analyzed decreases most often by more than an order of magnitude from one interval to the next as $\theta$ increases.  
Most pairs of publications that are co-cited at least ten times, therefore, have small values of $\theta$. 

Figure \ref{fig:mondrian} indicates when the null hypothesis of a best-fit lognormal or power law fitting the observed data can not be rejected.  We computed two types of statistics for evaluating the null hypothesis (${\chi}^2$ and K-S) and, moreover, we computed the ${\chi}^2$ statistic for four binning strategies. Figure \ref{fig:mondrian} indicates a distributional fit, specifically, if either the K-S $p$-value is greater than 0.05 or if two or more of the ${\chi}^2$ statistics are greater than 0.05.  
While we computed the K-L Divergence (see supplementary material), we did not use these computations for formal statements of distributional fit because they are neither a norm nor do they determine statistical significance. 
These K-L computations did, however, support the findings based on formal tests of statistical significance.

Power law distributions fit most often when co-citations are connected (Fig.~\ref{fig:mondrian}), when more extreme right tails are considered, and when co-citations have small values of $\theta$.  Lognormal distributions fit, conversely, in some circumstances, when a greater portion of the right tail is considered.  These observations support the existence of heavy tails for $\theta$ small, even if a lognormal distribution fits the observed data more broadly.  This observation is consistent with our observations of the most frequent co-citations having small $\theta$ values, as shown in the scatter plot in the left panel of Figure \ref{fig:Figs_4_5}. 

Mitzenmacher \cite{Mitzenmacher_2003} shows a close relationship between the power law and lognormal distributions vis-\`a-vis subtle variations in generative mechanisms that determine whether the resulting distribution is power law or lognormal. 
The stratified layers in Figure \ref{fig:mondrian} where a lognormal distribution fits for some portion of the right tail and, in the same instance, a power law describes the more extreme tail, 
may, therefore, be due to a generative mechanism whose parameters are close to those for a power law distribution as well as those for a lognormal distribution.

\begin{table}[!ht]
\caption{{\bf Exponents of best-fit power law distributions} These observations are for power law exponents where comparison across intervals of $\theta$  were possible, and where statistical tests indicated that a power law was a good fit to the data.  The articles of the co-citations were connected for all data shown.} 
\label{tab:exponents}
\centering
\begin{tabular}{ccc}
\hline
Right-tail cutoff ($\underline{x}$)  & $\theta$ & Power law exponent ($\alpha$) \\
\hline
200 & $\left[ 0.0 , 0.2 \right)$   & 3.26  \\
200 & $\left[ 0.2 , 0.4 \right)$   & 3.37  \\ \hline
250 & $\left[ 0.0 , 0.2 \right)$   & 3.27  \\
250 & $\left[ 0.2 , 0.4 \right)$   & 3.37  \\ \hline
300 & $\left[ 0.0 , 0.2 \right)$   & 3.22  \\
300 & $\left[ 0.2 , 0.4 \right)$   & 3.35  \\
\hline
\end{tabular}
\end{table}

\begin{figure}[h]
\includegraphics[width=0.5\textwidth]{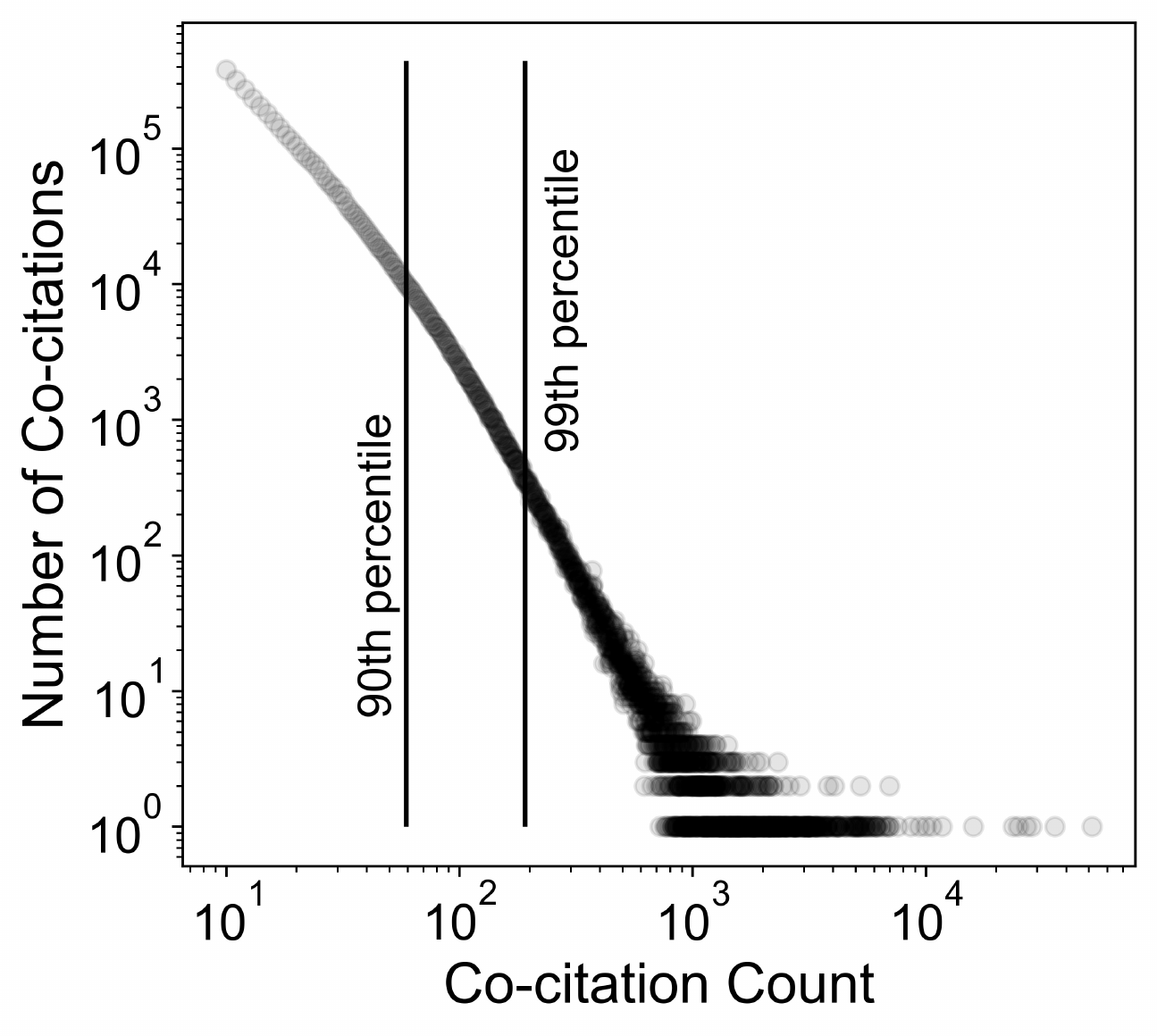}
\caption{\textbf{Log-log plot of the number of co-citations versus co-citation count for $\theta \in \left\lbrack 0.0, 0.2 \right)$} The $y$-axis shows the number of co-cited pairs observed having the citation counts plotted along 
the $x$-axis. The tightly clustered plot below the 99$th$ percentile demonstrates a clear pattern of decreasing number  of co-cited pairs having an increasing number of citation counts.  The scatter plot for the tail above 
the 99$th$ percentile broadens, indicating a long tail of relatively few co-cited pairs that were cited with extreme frequency.}
\label{fig:perc}
\end{figure}

Table 2 shows the exponents of the best-fit power law distributions when statistical tests indicated that a power law was a good fit and where comparisons were possible among the intervals of $\theta$: these were possible for  
$\theta$ intervals of $\left[ 0.0 , 0.2 \right)$ and $\left[ 0.2 , 0.4 \right)$, for connected co-citations, and right tails commencing at $\underline{x} \in \left\lbrace 200,250,300 \right\rbrace$. The power law exponent $\alpha$ in these comparisons  was less for $\theta \in \left[ 0.0 , 0.2 \right)$ than for $\theta \in \left[ 0.2 , 0.4 \right)$, indicating heavier tails for $\theta$ small and, therefore, a greater chance of extreme co-citation frequency.  
Figure \ref{fig:perc} shows a log-log plot of the number of co-citations ($y$-axis) exhibiting the counts on the $x$-axis, for $\theta$ in the interval $\left[0.0, 0.2 \right)$ (note that both axes employ log scaling).  The pattern for points below the 99th percentile clearly indicate that the number of co-citations referenced at a given frequency decreases greatly as the frequency increases.  Also, the broadening of the scatter where fewer co-citations are cited more frequently is indicative of a long right tail, as has been observed in other research where lognormal or power law distributions have been fit to data, as in \cite{montebruno_tale_2019}.  

Perline~\cite{perline_2005} warns against fitting a power law function to truncated data. Informally, a portion of the entire data set can appear linear on a log-log plot, while the entire data set would not. He cites instances where researchers have mistakenly characterized an entire data set as following a power law due to an analysis of only a portion of the data, when a lognormal distribution might provide a better fit to the entire data set.  Indeed, the scatter plot in Figure \ref{fig:perc} is not linear and so, as Figure \ref{fig:mondrian} shows, a power law does not fit the entire data set.  This is what Perline calls a weak power law where a power law distribution function fits the tail, but not the entire distribution. Our concern, however, is not with characterizing the distributional function for the entire data set, but with characterizing the features of high frequency co-citations, which by definition means we
were concerned with the right tail of the distribution.  Moreover, the results avoid confusion between lognormal and power law distribution functions because we have shown not only that a power law provides a statistically significant fit, but also  that a lognormal distribution function does not fit.

Our analysis found particularly heavy tails that were well fit by power law distributions for small $\theta$, in the intervals $\left[ 0.0, 0.2 \right)$ and $\left[ 0.2, 0.4 \right)$, and for co-citations whose constituents are connected, as shown in Fig.~\ref{fig:mondrian}. 
The closely related Matthew Effect~\cite{merton_1968}, cumulative advantage~\cite{price_general_1976}, and the preferential attachment class of models~\cite{barabasi_albert_2002} provide a possible explanation for citation frequencies following a power law distribution for some sufficiently extreme portion of the right tail. For greater values of $\theta$, insufficient data in the right tails precludes a definitive assessment in this regard, although one might argue that the lack of observations in the tails is counter to the existence of a power law relationship. It is also noteworthy that the exponents we found for co-citations (Table \ref{tab:exponents}) are close in value to those reported for citations by \cite{price_general_1976} and \cite{radicchi_statistical_2008}.  

\emph{Delayed Co-citations} The delayed onset of citations to a well cited publication, also referred to as `Delayed Recognition' and 'Sleeping Beauty', has been studied by Garfield, van Raan, and others~\cite{garfield_1970,vanraan_2019,ke_defining_2015,vanraansleeping2004,li_distinguishing_2016,glanzel_myth_2004,bornmann_identifying_2018}. 
We sought to extend this concept to frequently co-cited articles. As an initial step, we calculated two parameters (Materials and Methods): (1) the beauty coefficient~\cite{ke_defining_2015} modified for 
co-cited articles and  (2) timelag $t_l$, the length of time between first possible year of co-citation and the first year in which a co-citation was recorded. We further focused our consideration of delayed co-citations 
to the 95th percentile or greater of co-citation frequencies in our dataset of 4.1 million co-cited pairs. Within the bounds of this restriction, 24 co-cited pairs have a beauty coefficient of 1,000 or greater and all 24 are 
in the 99th percentile of  co-citation frequencies. Thus, very high beauty coefficients are associated with high co-citation frequencies.

\begin{figure}[h]
\includegraphics[width=0.8\textwidth]{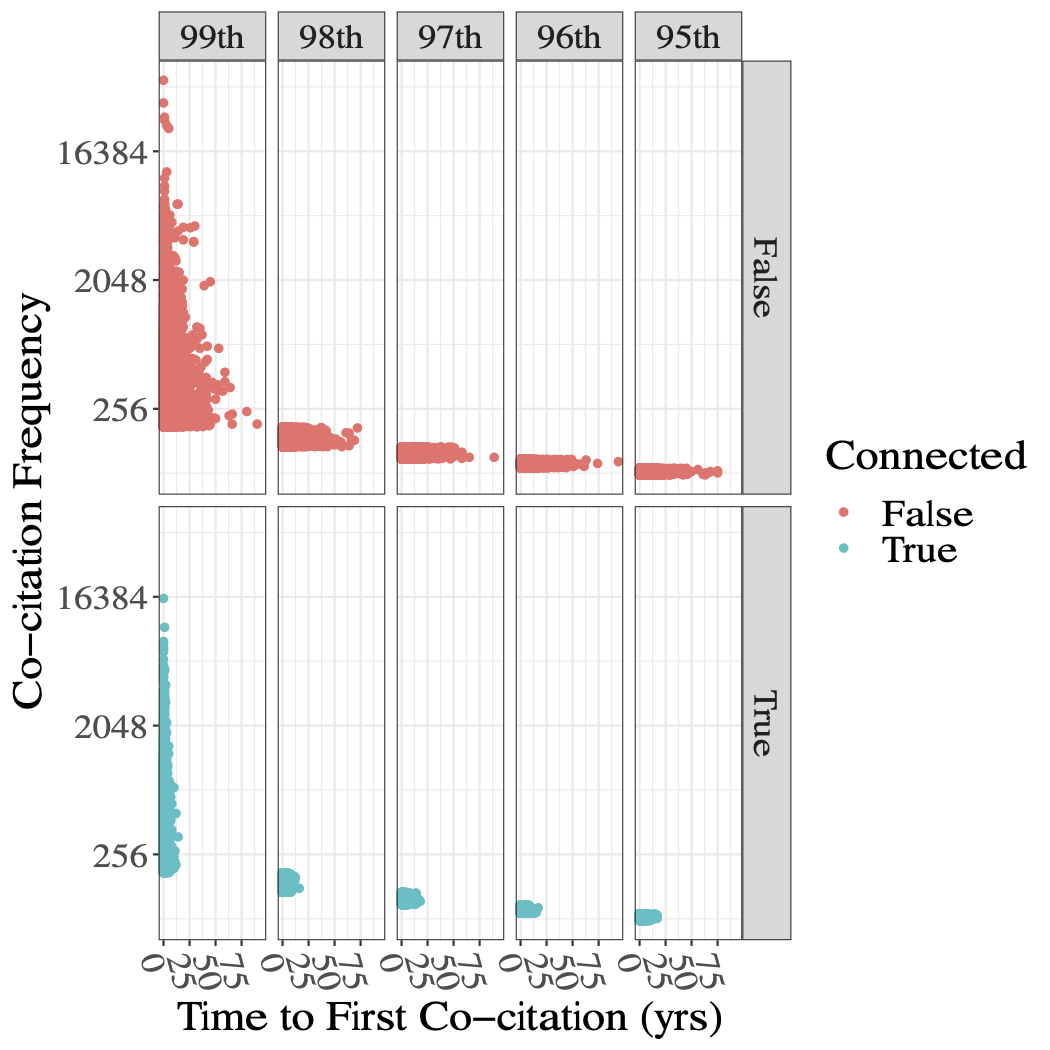}
\caption{\textbf{Relationship between time lag ($t_l$) and co-citation frequency} Extended lag times are associated with lower co-citation frequencies. Connected pairs have lower $t_l$ values. Data are shown for 207,214 pairs 
consisting of $\geq$  95th percentile of co-citation frequencies for the 4.1 million row dataset. The observations are stratified by percentile group (vertical panels) and connectedness (upper and lower halves). Co-citation
 frequency (y-axis) is plotted against  $t_l$, the time between first possible co-citation and first co-citation.}
\label{fig:composite}
\end{figure}

We also examined the relationship of $t_l$ with co-citation frequencies (Fig.~\ref{fig:composite}) and observed that high $t_l$ values were associated with lower co-citation frequencies. These data in appear to be consistent with a report from van Raan and Winnink~\cite{vanraan_2019}, who conclude that `probability of awakening after a period of deep sleep is becoming rapidly smaller for longer sleeping periods'. Further, when two articles are connected, they tend to have smaller $t_l$ values compared to pairs that are not connected in the same frequency range. 
\clearpage
\section*{Figures}

\clearpage

\clearpage

\clearpage

\begin{figure}[h]
\includegraphics[width=0.7\textwidth]{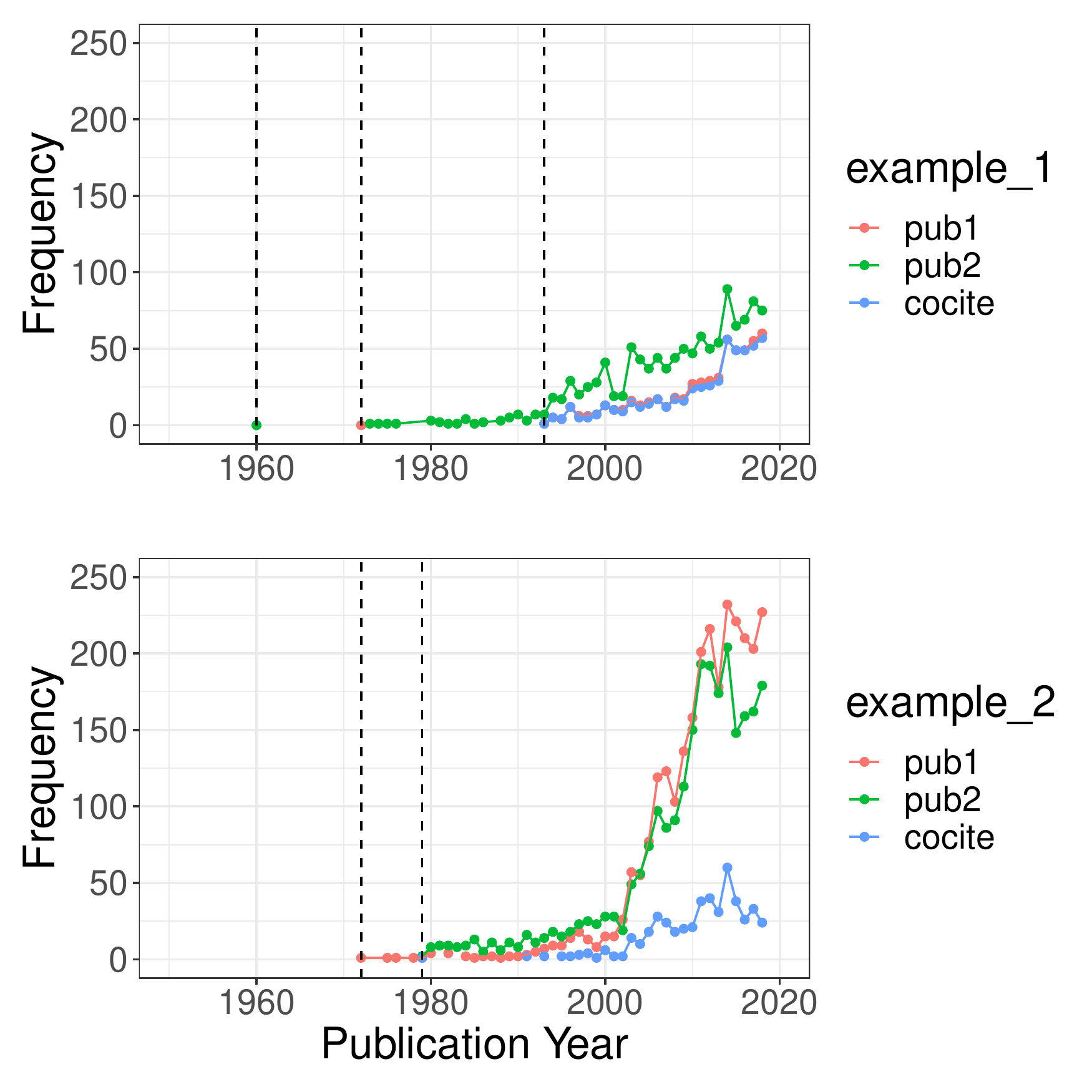}
\caption{\textbf{Co-citation frequencies of highly cited publications from Scopus 1985-1995} \emph{Upper panel} Publication 1: Instability of the interface of two gases accelerated by a shock wave 
(1972) doi: 10.1007/BF01015969, first cited (1993), total citations (566). Publication 2: Taylor instability in shock acceleration of compressible fluids (1960) doi: 10.1002/cpa.3160130207, first cited (1973), 
total citations (566), first co-cited (1993), total co-citations (541).
\newline 
\newline
\emph{Lower Panel} Publication 1: Colorimetric assay of catalase doi: 10.1016/0003-2697(72)90132-7 (1972) doi: 10.1016/0304-4165(79)90289-7, first cited (1972), total citations (2683). Publication 2: Levels 
of glutathione, glutathione reductase and glutathione S-transferase activities in rat lung and liver (1979) doi: 10.1016/0304-4165(79)90289-7, first cited (1979), total citations (2464), first co-cited (1979), total co-citations (470).} 
\label{fig:composite2}.
\end{figure}
\clearpage

\section*{Conclusions}

In this article, we report on our exploration of features that impact the frequency of co-citations. In particular, we wished to examine article pairs with high co-citation frequencies with respect to whether they originated from the same school(s) of thought or represented novel combinations of existing ideas. However, defining a discipline is challenging, and determining the discipline(s) relevant to specific publications remains a challenging problem. 
Journal-level classifications of disciplines have known limitations and while article-level approaches offer some advantages, they are not free of their own limitations~\cite{article_stasa}. 

Consequently, we designed $\theta$, a statistic that examines the citation neighborhood of a pair of articles $x$ and $y$ to estimate the probability that they would be co-cited. Our approach  has  advantages compared to alternate approaches: it avoids the challenges of journal-level analyses, it does not require a definition of ``discipline" (or ``disciplinary distance"), it does not require assignment of disciplines to articles,  it is computationally feasible, and, most importantly, it enables an evaluation that is specific to a given pair of articles.

We note that when  $x$ and $y$ are from the same sub-field, then $\theta$ may be very large, and conversely, when $x$ and $y$ are from very different fields, it might be reasonable to expect that $\theta$ will be small. Thus, in a 
sense, $\theta$ may correlate with disciplinary similarity, with large values for $\theta$ reflecting conditions where the two publications are in the same (or very close) sub-disciplines, and small values for $\theta$ reflecting that the 
disciplines for the two publications are very distantly related. We also comment that in this initial study, we have not considered second-degree information, that is publications that cite publications that cite an article of interest.

Our data indicate that the most frequent co-citations occur when co-citations have small values of $\theta$, as shown in Figure \ref{fig:Figs_4_5}. Our study considered the hypothesis that the 
frequency distribution is independent of $\theta$, but our statistical tests rejected this hypothesis, and 
showed instead that the frequency distribution is  best characterized by a power law for small values of $\theta$ and connected publications, and in many other regions is best characterized by a lognormal distribution.

The observation that power laws are consistent with small values of $\theta$ and connected co-citations is consistent with the theory of preferential attachment for these parameter settings. To the extent that 
preferential attachment is the mechanism giving rise to a power law, this suggests that preferential attachment is, at least, stronger for small $\theta$ values and connected co-citations than for other parameter combinations, 
or that preferential attachment is not applicable to other parameter values.

Observing power laws, heavy tails, and pairs with extreme co-citation strength for small values of $\theta$ (i.e.,  pairs that have small {\em a priori} probabilities of being co-cited) may seem, on its face, paradoxical. 
One possible explanation of the pairs in the extreme right tail with both small $\theta$ and large co-citation strength is that those pairs represent novel combinations of ideas that, when recognized within the research community, catalyze an 
increased citation rate, consistent with preferential attachment coupled to time-dependent initial attractiveness~\cite{eom_2011} as an underlying generative mechanism.  However, small values of $\theta$ do not guarantee 
a high co-citation count: indeed, even for small values of $\theta$, co-citations with a power law predominantly have relatively low co-citation strength. 

We also note the increasing proportion of connected pairs as the percentile for co-citation frequency
increases (Fig. \ref{fig:basedata}); this pair of parameters appears to be associated with a fertile environment where extremely high co-citation frequencies are possible.
This  observation raises the question of whether small values of $\theta$ and connected co-citations are associated with preferential attachment and, if a causal relationship exists, then 
how do $\theta$ and co-citation connection provide an environment supporting preferential attachment? A possibility is that one article in a co-cited pair citing the other makes the 
potential significance of the combination of their ideas apparent to researchers. The clear pattern of the highest frequency co-cited pairs  typically having low $\theta$ values suggests  that these pairs 
are highly cited and hence impactful because of the novelty in the ideas or fields that are combined 
(as reflected in low  $\theta$). However, other factors should be considered,  such as the prominence of authors and prestige of a journal~\cite{garfield_1980} where the first co-citation appears.

We did not apply field-normalization techniques when assembling the parent pool of 768,993 highly cited articles consisting of the top 1\% of highly cited articles from each year in the Scopus bibliography. 
Thus, the highly co-cited pairs we observe are biased towards high-referencing areas such as biomedicine and parts of the physical sciences~\cite{small_citation_1980}.
However, the dataset we analyzed has a lower bound of 10 on co-citation frequencies and includes pairs from fields other than those that are high referencing. 
For example, the maximum  $t_l$ we observed in the dataset of 4.1 million pairs was 149 years,
and is associated to a pair of articles  independently published in 1840, establishing their eponymous  
Staudt-Clausen theorem \cite{clausen_1840,vonstaudt_1840}; this pair of articles was apparently co-cited 10 times since their publication. 
A second pair of articles  concerning electron theory of metals~\cite{drude_1900_1,drude_1900_2} was first co-cited in 1994 for a total of 109 times, with $t_l$ observed of 94 years. Both cases are drawn from mathematics and physics rather than the medical literature. They are also consistent with the suggestion that the probability of awakening is smaller after a period of deep sleep~\cite{vanraan_2019}. 
As we have defined $t_l$, with its heavy penalty for early citation, we create additional sensitivity to coverage and data quality especially for pairs with low citation numbers. Indeed, for the  Staudt-Clausen  pair, a manual search of other sources revealed an article~\cite{carlitz_cvs_1961} in which they are co-cited. Both these articles were originally published in German and it is possible that additional co-citations were not captured. Thus, big data approaches that serve to identify trends should be accompanied by more meticulous  case studies, where possible. Other approaches for examining depth of sleep and awakening time should certainly be considered~\cite{vanraansleeping2004,ke_defining_2015}. 
Lastly, using our approach to revisit invisible colleges~\cite{price_beaver_1966,crane_1972,small_clustering_1985} seems warranted, since it seems likely that the upper bound of a hundred members predicted 
by \cite{price_beaver_1966}
 is likely to have increased in a global scientific enterprise with electronic publishing and social media. 

Finally, we view these results as a first step towards further investigation of co-citation behavior, and we introduce a new technique based on exploring first-degree neighbors of co-cited publications; we are hopeful that this  graph-theoretic study will stimulate new approaches that will provide additional insights, and prove complementary to other article level approaches. 

\section*{Acknowledgments}  In addition to support through federal funding, the ERNIE project features a collaboration with Elsevier.  We thank our colleagues from Elsevier for their support of the collaboration.

\section*{Competing Interests}

The authors have no competing interests. Scopus data used in this study was available to us through a collaborative agreement with Elsevier on the ERNIE project. Elsevier personnel played no role in conceptualization, experimental design, review of results, or conclusions presented. The content of this publication is solely the responsibility of the authors and does not necessarily represent the official views of the National Institutes of Health or Elsevier. Sitaram Devarakonda's present affiliation is Randstad USA. His contributions to this article were made while he was a full-time employee of NET ESolutions Corporation.

\section*{Author Contributions}
Conceptualization, GC, JB, SD, and TW; Methodology, AD, DK, GC, JB, SD, SL, and TW; Investigation, DL-H, GC, JB, and SD; Writing -Original Draft, GC, JB, TW; Writing- Review and Editing, AD, DK, DL-H, GC, JB, SD, SL, and  TW; Funding Acquisition, GC; Resources, DK and GC; Supervision, GC. Authors are listed in alphabetic order.

\bibliographystyle{acm}
\bibliography{R1arxiv.bib}
\end{document}